\documentclass[aps,twocolumn,nofootinbib,showpacs, prd]{revtex4-2}

\usepackage{hyperref}
\usepackage{amsmath,amsfonts,amssymb}
\hypersetup{dvips,dvipdfm,colorlinks=true,urlcolor=magenta,filecolor=magenta,linktoc=page,citecolor=red,linkcolor=blue,bookmarks=true}
\usepackage{graphicx,epstopdf}
\usepackage{multirow}
\usepackage{natbib}
\begin{document}
	\title{Curvature form of Raychaudhuri equation and its consequences: A geometric approach}
	\author{Madhukrishna Chakraborty} \email{chakmadhu1997@gmail.com}
	\author{Subenoy Chakraborty}\email{schakraborty.math@gmail.com (corresponding author)}
	\affiliation{Department of Mathematics, Jadavpur University, Kolkata - 700032, India}
	\begin{abstract}
	The paper aims at deriving a curvature form of the famous Raychaudhuri equation (RE) and the associated criteria for focusing of a hyper-surface orthogonal congruence of time-like geodesic. Moreover, the paper identifies a transformation of variable related to the metric scalar of the hyper-surface which converts the first order RE into a second order differential equation that resembles an equation of a Harmonic oscillator and also gives a first integral that yields the analytic solution of the RE and Lagrangian of the dynamical system representing the congruence.
	\end{abstract}
	\maketitle
	\small ~Keywords :  Raychaudhuri Equation; Focusing; Extrinsic curvature; Intrinsic curvature;  Harmonic oscillator ; First Integral; Lagrangian formulation.
	\section{Introduction}
	Raychaudhuri equation (RE) is a very important and useful equation in understanding the dynamics of our universe and hence finds an immediate application in General Relativity (GR), Cosmology and Astrophysics \cite{Wald:1984rg}, \cite{Weinberg:1972kfs}. However, being an identity in Lorentzian geometry, the scope of RE is not restricted to these areas but beyond it. Precisely speaking, the equation encodes information regarding ``\textit{flows}". Flows are generated by vector fields. The corresponding integral curves of these vector fields in turn identify the flow. The nature of these curves may be geodesic or non-geodesic. However, in the context of gravity theory the former is more important. In GR \cite{Weinberg:1972kfs}, RE implies Focusing theorem (FT)/ the notion of geodesic focusing which marks the inevitable existence of singularity in Einstein gravity \cite{Raychaudhuri:1953yv}. The condition on matter to facilitate focusing is termed as Convergence Condition (CC). This CC is equivalent to the positivity of the Raychaudhuri scalar \cite{Kar:2006ms}. Penrose used the notion of geodesic focusing and geodesic incompleteness to prove the seminal singularity theorems in GR \cite{Penrose:1964wq}, \cite{Hawking:1970zqf}. The singularity theorems are believed to be the greatest accomplishment of GR \cite{Hawking:1970zqf}. It is interesting to note that the RE is seed of the singularity theorems. Singularity means a condition where the density becomes infinite or equivalently a situation of zero volume. However, a formal geometric definition of what a singularity is first appeared in Penrose's work \cite{Penrose:1964wq}. The notion of singularity and geodesic incompleteness were aligned by him \cite{Senovilla:2022vlr}, \cite{Senovilla:2021pdg}, \cite{Senovilla:2014gza}. Although, GR is the most well acclaimed theory of gravity that supports a plethora of physical phenomena yet the occurrence of singularities in Einstein's theory is the biggest drawback of the theory \cite{Landsman:2022hrn}. It is generally speculated that since in modified theories of gravity the field equations differ from that of GR, there might arise some possibilities for the possible avoidance of focusing. Motivated by this, several authors studied the modified RE and corresponding CC in a plethora of extended theories of gravity with all possible geometric backgrounds like $f(R)$ gravity \cite{Chakraborty:2023ork}, $f(T) $ gravity \cite{Chakraborty:2023yyz}, scalar tensor theories \cite{Choudhury:2021zij}, Kantowski-Sachs model \cite{Chakraborty:2023rgb} etc. Several cosmological interpretation of RE and FT can be found in \cite{Chakraborty:2023lav}, \cite{Chakraborty:2024wty}. Moreover, there are some review articles that feature RE and its associated aspects, for details see ref \cite{Kar:2006ms}, \cite{Chakraborty:2024khs}. It is worthy to mention that, RE is not only important in hinting the singular nature of GR but also it gives possible resolution pathways for the same. In this context, a quantum analogue of classical RE and quantum corrected RE deserve special mention. RE in quantum settings have been attempted in \cite{Chakraborty:2023voy}-\cite{Blanchette:2020kkk}. The derivation of Raychaudhuri equation is somewhat different from the way it is presented in standard textbooks for a general readership \cite{Dadhich:2005qr}. Raychaudhuri however in subsequent works deduced the equation in modern approach. His motivation of deriving the equation was entirely restricted to cosmology. He assumed universe to be characterized by a time-dependent geometry but did not assume homogeneity and isotropy at the outset. His main aim was to distinguish the role of spin (non-zero rotation/ vorticity) and shear (anisotropy) in formation and/ or avoidance of singularity \cite{Bhatt:2021hdi}. Raychaudhuri equation as mentioned before is a geometric identity. The last term on the r.h.s of the equation (involving the (0,2) Ricci tensor) encodes information regarding any particular gravity theory. In 1955, Heckmann and Schuking \cite{book1} got a set of equations in the context of Newtonian cosmology. One of the equations resemble RE. Raychaudhuri was prompted by this work and re-derived the same set of equations in a different manner in a subsequent article. In 1961, Jordan et.al \cite{book2} extensively wrote an article on the relativistic mechanics of continuous media. In this article, evolution of the other kinematic quantities namely shear and rotation appeared for the first time in literature. RE for null geodesic was derived by Sachs \cite{Sachs:1962wk}.   RE is often referred to as Landau-Raychaudhuri equation \cite{book3} because Landau had significant contribution in proving the fact that focusing alone does not lead to the formation of space-time singularities but the converse is true. That means, if there is a singularity, focusing will inevitably occur there. This is a consequence of attractive gravity. This seemingly trivial concept of geodesic focusing due to attractive gravity is shown by FT. That is why RE is often called the fundamental equation for gravitational attraction. The present work is an attempt to derive the RE in terms of different notion of curvature and consequently, the name ``curvature form" has been coined. This is because, Einstein gave the revolutionary idea that curvature of space-time is due to the attractive nature of gravity. Again, RE hints the inevitable existence of singularity in Einstein gravity via the notion of geodesic focusing and this is also a consequence of attractive gravity. To be precise, a space-time singularity or a gravitational singularity is a situation where gravity is assumed to be so intense that the space-time itself would break catastrophically (divergence/ blowing off curvature). Since RE studies singularity via behavior of geodesics. This motivates us to derive a curvature RE to see the effect of curvature in focusing/defocusing of geodesics. Moreover, with an aim to study the role of extrinsic and intrinsic curvature in focusing we have attempted a deduction of curvature form of RE and FT. The motivation is purely from the point of view of geometry and in search for some geometries where focusing might be avoided leading to possible avoidance of singularity. Subsequently, a transformation of variable is considered which converts the first order RE to a second order differential equation analogous to the evolution equation of a classical harmonic oscillator. In literature, Harmonic oscillator form of RE is very popular \cite{Kar:2006ms}. Actually, such a form argues that if RE corresponds to a real Harmonic oscillator then focusing is inevitable and the focusing scalar is associated with the time varying frequency of the oscillator. In the present work, the transition to harmonic oscillator form has been done to see the effect of curvature in frequency of the oscillator and in this context a notion of curvature frequency has been introduced. Moreover, the analytic solution of the RE has been attempted by first integral formulation.
 The layout of the paper is as follows: Section II discusses the original form of the RE and geodesic Focusing theorem. Section III deals with the derivation of RE in curvature form and deduction of criterion for focusing. Section IV shows the Harmonic oscillator analysis and first integral formulation. The paper ends with concluding remarks in Section V.
 \section{Raychaudhuri equation and Focusing theorem: An overview}
  The RE deals with the kinematics of flows \cite{Kar:2006ms}. Although  RE is generally true for any curves that may be time-like or null, its geodesic version can be deduced from the general form very easily for, in the context of gravity the geodesic version is more important and relevant.
 Let us now introduce a formal definition of a \textit{congruence}. Let $\mathcal{M}$ be a manifold and $U$ be open in $\mathcal{M}$. A \textit{congruence} is defined as a family of curves in $U$ if precisely one curve in this family passes through each point $p\in U$. Now, we focus our attention to study the kinematic quantities associated with such flows and how  Raychaudhuri equation is related to them \cite{Horwitz:2021lyc}. In this context, it is to be noted that evolution equation of the kinematic quantities (which characterize the flow) in a given space-time background along the flow is of central importance.
 Let $\tau$ be the parameter labeling points on the curves and $v^{a}$ be the velocity vector field along the congruence. Let,  $R_{ab}$ be the $(0,2)$ Ricci tensor projected along the geodesic flow. Essentially one needs to have various functions of $\tau$ in order to characterize the flow. To define the kinematics of a deformable medium, we consider the gradient of velocity vector field which can be geometrically represented by a $(0,2)$ deformation tensor, say $\mathcal{B}_{ab}$. Further, this tensor can be split into four fundamental and irreducible tensors as follows:
 \begin{equation}
 	\mathcal{B}_{ab}=\nabla_{b}v_{a}=\dfrac{\Theta}{n-1}~ \eta_{ab}+\sigma_{ab}\sigma^{ab}+\omega_{ab}\omega^{ab}-A_{a}v_{b}\label{eq01},
 \end{equation} where, $\nabla$ denotes the \textit{covariant derivative} and $n$ is the dimension of the space-time. One may check that $	\mathcal{B}_{ab}v^{b}=\mathcal{B}_{ab}v^{a}=0$ as $v^{a}$ satisfies the geodesic equation and is normalized as $v^{a}v_{a}=-1$. Now, we discuss the geometric and physical meaning of the introduced tensors in eq.(\ref{eq01}).
 \begin{itemize}
 	\item To analyze the physical interpretation of $\mathcal{B}_{ab}$, a one parameter subfamily of geodesics in the congruence denoted by $\gamma_{l}(\tau)$ needs to be considered. Here $l$ is the labeling parameter. The geodesics in the congruence can be collectively represented by $x^{a}(\tau, l)$. Let $\xi^{a}$ be the orthogonal deviation vector from a reference geodesic say $\gamma_{0}$ in the subfamily. Using $v^{a}=\dfrac{\partial x^{a}}{\partial \tau}$ and $\xi^{a}=\dfrac{\partial x^{a}}{\partial l}$ we have 
 	\begin{equation}
 		v^{b}\nabla_{b}\xi^{a}=\xi^{b}\nabla_{b}v^{a}=\mathcal{B}^{a}_{b}\xi^{b}.
 	\end{equation}
 	$\mathcal{B}^{a}_{b}$ measures the failure of $\xi^{a}$ to be transported parallelly. The symmetry of $\mathcal{B}_{ab}$ i.e, $\mathcal{B}_{ab}=\mathcal{B}_{ba}$ is ensured by the Frobenius' theorem.
 	\item \textbf{$\Theta$} is known as the expansion scalar. It is the trace part of $	\mathcal{B}_{ab}$ i.e, $\Theta=\mathcal{B}^{a}_{a}=\nabla_{a}v^{a}$. It describes the average separation between the geodesic worldlines of the $v_{a}$--congruence, precisely the average expansion/contraction of the associated observers. This means, if one considers the congruence as a collection of flow lines (geodesics) then cross sectional area enclosing the geodesics may evolve along the congruence. If the geodesics go apart or come closer then area will increase or decrease accordingly. This phenomenon is demonstrated by $\Theta$, the expansion scalar. 
 	\item $\sigma_{ab}$ is known as the shear tensor defined by $\sigma_{ab}=\dfrac{1}{2}\left(\nabla_{b}v_{a}+\nabla_{a}v_{b}\right)-\dfrac{\Theta}{n-1}\eta_{ab}$. It is symmetric traceless part of $	\mathcal{B}_{ab}$ i.e, $\sigma_{ab}=\sigma_{ba}$ and $\sigma^{a}_{a}=0$. It measures the kinematic anisotropies. The shape of the area enclosing congruence of geodesics may be sheared and this is demonstrated by $\sigma$.
 	\item $\omega_{ab}$ is the anti-symmetric part of $	\mathcal{B}_{ab}$ (i.e, $\omega_{ab}=-\omega_{ba}$) and is defined by $\omega_{ab}=\dfrac{1}{2}\left(\nabla_{b}v_{a}-\nabla_{a}v_{b}\right)$. It is called the rotation/ vorticity tensor as it measures the kinematic rotation or monitors the rotational behavior of the $v_{a}$--vector field.  The shape of the area enclosing the congruence of geodesics may be twisted. This is demonstrated by $\omega_{ab}$. A congruence is hyper-surface orthogonal if and only if $\omega_{ab}=0$. This is a consequence of the Frobenius' theorem discussed in the previous subsections in the context of differential geometry.
 	\item $\eta_{ab}=g_{ab}\pm v_{a}v_{b}$ is called the induced metric/projection tensor that operates on the $(n-1)$ dimensional hyper-surface. $\eta_{ab}$ satisfies the orthogonality condition i.e, $v^{b}\eta_{ab}=0$. In expression for $\eta_{ab}$, `+' sign is for time-like curves ($v_{a}v^{a}=-1$) and `-' sign is for null curves ($v_{a}v^{a}=0$).
 	\item $A_{a}$ is the 4-acceleration vector field defined by $A_{a}=v^{b}\nabla_{b}v_{a}$. This field guarantees the presence of non gravitational forces. Therefore, $A_{a}=0$ in case of geodesic worldlines.
 \end{itemize}
 Thus, the above discussion shows that the expansion, rotation and shear are purely geometric characteristic of the cross sectional area enclosing a bundle of curves orthogonal to the flow lines. The shape of this area changes as one moves from one point to another along the flow but it still includes the same set of curves in the bundle. One thing which may change during the flow is that the bundle may be isotropically smaller or larger, sheared or twisted. This situation seems to be analogous to the elastic deformations or fluid flow.
 The general Raychaudhuri equation \cite{Burger:2018hpz}, \cite{Bhattacharyya:2021djv} for non-geodesic motion is nothing but the proper time ($\tau$) evolution of the expansion scalar ($\Theta$)  as
 \begin{equation}
 	\dfrac{d\Theta}{d\tau}=-\dfrac{\Theta^{2}}{n-1}-\sigma_{ab}\sigma^{ab}+\omega_{ab}\omega^{ab}+\nabla_{b}A^{b}-R_{ab}v^{a}v^{b}\label{eq02**}
 \end{equation}
 Raychaudhuri equation for null geodesic congruence is given by
 \begin{equation}
 	\dfrac{d\Theta}{d\lambda}=-\dfrac{\Theta^{2}}{n-2}-\sigma_{ab}\sigma^{ab}+\omega_{ab}\omega^{ab}+\nabla_{b}A^{b}-R_{ab}k^{a}k^{b}\label{eq002*}
 \end{equation} where $k^{a}$ is a null vector and $\lambda$ is an affine parameter.
 Ricci tensor $R_{ab}$ is a $(0,2)$ tensor that carries the effect of local gravitational field as otherwise RE is a purely geometric identity and has nothing to do with gravity. The term $-R_{ab}v^{a}v^{b}$ encapsulates the contribution of space-time geometry and is independent of the derivatives of the vector field. Thus in comparison to the other terms present in eq. (\ref{eq02**}) and (\ref{eq002*}), this particular term possesses more general implications. It has a geometrical interpretation as a mean curvature in the direction of the $v_{a}$--congruence \cite{Albareti:2014dxa}, \cite{Albareti:2012se}. This idea is elaborated in the following sections.
The general form of the RE (\ref{eq02**}) can be reduced to much simpler form if one assumes:\\
1. The congruence of curves to be time-like geodesic. In that case $A^{b}=0$.\\
2. Congruence of time-like geodesics to be hyper surface orthogonal, which by virtue of Frobenius theorem \cite{Poisson:2009pwt} of differential geometry implies zero rotation i.e, $\omega_{ab}=0$. Since $\sigma_{ab}$ is spatial hence, $\sigma_{ab}\sigma^{ab}\geq0$. If one is interested in geodesic focusing then zero vorticity congruence must be taken into consideration to avoid centrifugal forces.\\
Thus the simplified version of the original equation (\ref{eq02**}) upon assuming the above conditions is given by 
\begin{equation}
	\dfrac{d\Theta}{d\tau}=-\left(\dfrac{\Theta^{2}}{n-1}+2\sigma^{2}+\tilde{R}\right)\label{eq03*}
\end{equation} where $2\sigma^{2}=\sigma_{ab}\sigma^{ab}$ and $\tilde{R}=R_{ab}v^{a}v^{b}$ (Raychaudhuri scalar). In addition if $\tilde{R}\geq0$ then $\dfrac{d\Theta}{d\tau}\leq0$ which shows that the expansion gradually decreases with the evolution of the congruence. On an explicit manner we now analyze the Focusing Theorem as follows:\\
If matter content of the universe satisfies strong energy condition (SEC) i.e ,
\begin{equation}
	T_{ab}v^{a}v^{b}+\dfrac{1}{2}T\geq0 ,
\end{equation} where $T_{ab}$ is the energy momentum tensor
then Einstein's equation 
\begin{equation}\label{eq3}
	R_{ab}-\dfrac{1}{2}Rg_{ab}=T_{ab}
\end{equation}
yields, \begin{equation}\label{eq77*}
	R_{ab}v^{a}v^{b}\geq0.
\end{equation}
Employing the condition (\ref{eq77*}) on (\ref{eq03*}) we get ,
\begin{equation}
	\dfrac{d\Theta}{d\tau}+\dfrac{\Theta^{2}}{n-1}\leq0.
\end{equation}
Integrating the above inequality w.r.t proper time $\tau$ we get,
\begin{equation}
	\dfrac{1}{\Theta(\tau)}\geq\dfrac{1}{\Theta_{0}}+\dfrac{\tau}{n-1}.\label{eq1.119}
\end{equation}
Thus, one can infer that any initially converging hyper-surface orthogonal congruence of time-like geodesics must continue to converge within a finite value of the proper time $\tau\leq-(n-1)\Theta_{0}^{-1}$ which leads to crossing of geodesics and formation of a congruence singularity (may or may not be a curvature singularity). This is called the FT and the condition (\ref{eq77*}) is the corresponding CC. Further, it is to be noted that the SEC causes gravitation to be attractive and hence can't cause geodesic deviation, rather it increases the rate of convergence. Thus the FT inevitably proves the generic existence of singularity as a major drawback of Einstein gravity. FT proves the seemingly trivial statement that gravity is attractive which draws the geodesics closer and make them converge in some finite value of the affine parameter/ proper time. FT in terms of null geodesic can be derived in a similar manner by considering the RE in case of null geodesic congruence except there is a change in the inequality as $	\dfrac{d\Theta}{d\lambda}+\dfrac{\Theta^{2}}{n-2}\leq0$. However, it is to be noted that focusing leads to congruence singularity which may or may not be a space-time singularity. However with certain global assumptions related to causality in Lorentzian geometry, focusing might lead to cosmological or black-hole singularity. Thus, focusing alone can not lead to singularity but the converse is true. A possible resolution of the initial big-bang singularity or black-hole singularity may be done by violating the FT (characterized by negativity of the Raychaudhuri scalar).
\section{Raychaudhuri equation and Focusing theorem in curvature form}
We consider Bianchi cosmologies for which the space-time is homogeneous and anisotropic in nature. We use, $(3+1)$-decomposition of the $4D$ manifold, with homogeneous hyper-surface (characterized by the three space metric $\eta_{ab}$) and the extrinsic curvature $K_{ab}$. They are defined as \cite{Chakraborty:2001wv}
\begin{equation}
	\eta_{ab}=g_{ab}+v_{a}v_{b}
\end{equation}
and
 \begin{equation}
 	K_{ab}=\dfrac{1}{2N}\dfrac{\partial g_{ab}}{\partial t}=\dfrac{1}{3}K\eta_{ab}+\sigma_{ab}
	\end{equation} where $K=K_{ab}\eta^{ab}$ is the trace of the extrinsic curvature, $\sigma_{ab}$ is the shear of the time-like geodesic congruence orthogonal to the homogeneous hyper-surfaces. The Hamiltonian constraint can be written as \cite{Poisson:2009pwt}, \cite{Chakraborty:2001wv}
\begin{equation}
	K^{2}=\dfrac{3}{2}\sigma_{ab}\sigma^{ab}-\dfrac{3}{2}~^{(3)}R+3T_{d}(v)\label{eq13}
\end{equation} and
\begin{equation}
	\dot{K}=-\dfrac{1}{3}K^{2}-\sigma_{ab}\sigma^{ab}-T_{s}(v)\label{eq14}
\end{equation} is the RE.  Here $^{(3)}R$ is the scalar curvature of the homogeneous hyper-surface.
\begin{eqnarray}
	T_{d}(v)=T_{ab}v^{a}v^{b}=G_{ab}v^{a}v^{b}=\left(R_{ab}-\frac{1}{2}Rg_{ab}\right)v^{a}v^{b}\\\nonumber=R_{ab}v^{a}v^{b}+\dfrac{1}{2}R=-M_{c}+\dfrac{R}{2}
\end{eqnarray} $M_{c}=-R_{ab}v^{a}v^{b}$ is the mean curvature along the normal to the (space-like) hyper-surface i.e, mean curvature is directed along the time-like geodesic orthogonal to the hyper-surfaces \cite{Albareti:2014dxa}, \cite{Albareti:2012se}.  Clearly,
\begin{equation}
	T_{s}(v)=\left(T_{ab}-\frac{1}{2}Tg_{ab}\right)v^{a}v^{b}=R_{ab}v^{a}v^{b}=-M_{c}\label{eq16}
	\end{equation}
We assume $\kappa=1$, then equations (\ref{eq13}) and (\ref{eq14}) gives
\begin{equation}
	\dot{K}+K^{2}=R-^{(3)}R-M_{c}\label{eq19}
\end{equation}
This form of RE is fully characterized by Ricci scalar $R$, 3-space curvature ($^{(3)}R$) and the mean curvature $(M_{c})$. Extrinsic geometry is the description of curves, surfaces and generalizations viewed as embedding in a space of higher dimension than themselves. While Intrinsic geometry is a way of describing curvature without appeal to higher dimension. One may note that $R=^{(4)}R$ is an intrinsic curvature while $^{(3)}R$ is an extrinsic curvature. Alternatively, using the Gauss-Codazzi equation one can write the interrelation between the 4-space curvature and 3-space curvature as
\begin{equation}
	^{(4)}R=^{(3)}R-(K^{2}-K^{ab}K_{ab})-2(v^{a}_{;~b}~v^{b}-v^{a}v^{b}_{;~b})_{;a}\label{eq20}
\end{equation} where $R=^{(4)}R$ is the $4D$ Ricci scalar. Using (\ref{eq20}) in (\ref{eq19}) one may write the RE as
\begin{equation}
	\dot{K}+K^{2}=-(K^{2}-K^{ab}K_{ab})-2(v^{a}_{;~b}~v^{b}-v^{a}v^{b}_{;~b})_{;a}-M_{c}
\end{equation}
Here $^{(4)}R$ is an intrinsic curvature while $^{(3)}R$ is an extrinsic curvature. $K_{ab}$ is intimately related to the normal derivative of the metric tensor. It is concerned with the extrinsic aspects-the way in which the hyper-surface is embedded in the enveloping space-time manifold. $K=\eta^{ab}K_{ab}$ is equal to the expansion of a congruence of geodesics which intersect the hyper-surface orthogonally (so that their tangent vector is equal to $v^{a}$ on the hyper-surface). $K>0$ implies that the hyper-surface is convex i.e, the the congruence is diverging while, $K<0$ implies that the hyper-surface is concave and the congruence is converging. Gaussian curvature is intrinsic while the mean curvature $M_{c}$ is extrinsic in nature.
In order to find the conditions for focusing we consider the following cases and assume the fact that effective curvature is finite and bounded (it admits both lower and upper bounds).\\

\textbf{Case-I}: $^{(4)}R-^{(3)}R-M_{c}=0$ (zero curvature/curvatures nullify the effect of each other so that the whole contribution is zero) \\
Upon integrating the equation (\ref{eq19}), we have $\dfrac{1}{K}=\dfrac{1}{K_{0}}+t$, where $K_{0}$ is a constant of integration such that $K(t=t_{0})=K_{0}$ (initial expansion scalar). This shows that at $t=-\dfrac{1}{K_{0}}$, $K$ blows (i.e, $K=\pm\infty$). $K\rightarrow+\infty$ implies a complete divergence while $K\rightarrow-\infty$ implies a complete convergence. This leads us to the conclusion that if there is no effect of curvature terms (either they are zero individually or even if they are non-zero their combined contribution is zero) then singularity occurs in finite time either in past or future depending on the sign of initial expansion $K_{0}$. For convergence, if the initial expansion $K_{0}$ is negative then focusing occurs in finite time in the future.\\

\textbf{Case-II}: ($^{(4)}R-^{(3)}R-M_{c})>0$ and has an upper bound $\Lambda_{0}^{2}$ (say). Then (\ref{eq19}) implies the inequality
\begin{equation}
		\dot{K}+K^{2}\leq\Lambda_{0}^{2}
\end{equation} which upon integration yields
\begin{equation}
	K\leq \Lambda_{0}\tanh(\Lambda_{0}(t-t_{0}))
\end{equation} Since $\tanh\Lambda_{0}(t-t_{0})$ is always finite and bounded, so $K$ (expansion) is always finite. Thus, focusing does not occur and there is no singularity.\\

\textbf{Case-III}: ($^{(4)}R-^{(3)}R-M_{c})<0$ and has a lower bound say $-\mu_{0}^{2}$. Then from the RE (\ref{eq19}) we have
\begin{equation}
	\dot{K}+K^{2}>-\mu_{0}^{2}
\end{equation} which upon integration yields,
\begin{equation}
	K>\mu_{0}\tan\left(\mu_{0}(t_{0}-t)\right)
\end{equation}
Clearly, $K$ diverges at $t=t_{0}-\dfrac{\pi}{2\mu_{0}}$ (finite time). Therefore, in this case focusing occurs in finite time.\\

Based on the above three cases we infer that the effective curvature scalar ($^{(4)}R-^{(3)}R-M_{c}$) has a significant role in focusing of the congruence of time-like geodesic. Therefore we name this as ``\textbf{\textit{Focusing curvature}}" and denote it by $F_{c}$. The above results in terms of $F_{c}$ may be summarized as follows:
\begin{enumerate}
	\item For $F_{c}=0$, focusing occurs in finite time and whether the congruence will converge in the future or past is determined by the sign of initial expansion $K_{0}$.
	\item For $F_{c}>0$, there is no singularity as focusing is avoided.
	\item For $F_{c}<0$, focusing occurs in finite time.
\end{enumerate}
Now we deduce a beautiful analogy between geometry and physics of FT by using (\ref{eq14}). We have,
\begin{equation}
\dot{K}+\dfrac{1}{3}K^{2}=-\sigma_{ab}\sigma^{ab}+M_{c}\label{eq26}
\end{equation} where we have used $\kappa=1$ and equation (\ref{eq16}).
In this case as $M_{c}=-R_{ab}v^{a}v^{b}$ and $\sigma_{ab}\sigma^{ab}=2\sigma^{2}$, so correlating the geometry and physics we can say that Strong Energy Condition (SEC) on matter in GR ($R_{ab}v^{a}v^{b}\geq0$ or the Convergence Condition) is equivalent to negative mean curvature (extrinsic) and vice-versa which yields $\dot{K}+\dfrac{K^{2}}{3}\leq0$. Now, similarly by (\ref{eq1.119}) one can deduce focusing theorem. Thus, geometrically one may say that negative mean curvature of space-time manifold tend to focus an initially converging hyper-surface orthogonal congruence of time-like geodesic within a finite time. Equivalently, there is no focusing (hence no singularity) in spaces with positive mean curvature. This treatment may be regarded as the curvature form of FT which shows that there is a significant role of extrinsic curvature in focusing, an important finding of the present work.\\

Let us now study focusing in inflationary era. For this we modify equations (\ref{eq13}) and (\ref{eq14}) by introducing a self-interacting scalar field so that (\ref{eq13}) and (\ref{eq14}) now become
\begin{equation}
	K^{2}=3\left(\dfrac{1}{2}\dot{\phi}^{2}+V(\phi)\right)+\dfrac{3}{2}\sigma_{ab}\sigma^{ab}-\dfrac{3}{2}~^{(3)}R+3T_{d}(v)\label{eq13*}
\end{equation} and
\begin{equation}
	\dot{K}=\left(-\dot{\phi}^{2}+V(\phi)\right)-\dfrac{1}{3}K^{2}-\sigma_{ab}\sigma^{ab}-T_{s}(v)\label{eq14*}
\end{equation} 
 In inflationary scenario the potential energy dominates over the kinetic energy of the inflaton field i.e, $V(\phi)>>\dot{\phi}^{2}$ ($\dot{\phi}^{2}$ can be neglected compared to $V(\phi)$) and further during the slow roll inflation the potential behaves as a cosmological constant $\Lambda_{0}$ i.e, $V(\phi)\approx V_{0}=\Lambda_{0}$.
 So, from equation (\ref{eq14*}) one gets
\begin{equation}
	\dot{K}+K^{2}\leq \Lambda_{0}
\end{equation} which upon integration yields,
\begin{equation}
	K\leq\sqrt{3\Lambda_{0}}\tanh\left(\sqrt{\frac{\Lambda_{0}}{3}}(t-t_{0})\right)
\end{equation}
Thus $K$, the expansion rate approaches to $\sqrt{3\Lambda_{0}}$ exponentially over the time  scale $\sqrt{\frac{3}{\Lambda_{0}}}$. Further, using equation (\ref{eq13*}) one obtains \cite{Chakraborty:2001wv}
\begin{equation}
	\dfrac{3}{2}\sigma_{ab}\sigma^{ab}\leq K^{2}\leq 3\lambda_{0} \tanh^{2}\left(\sqrt{\frac{\Lambda_{0}}{3}}(t-t_{0})\right)
\end{equation}
Clearly, there is no singularity in the inflationary era as $K$ is finite and bounded. This is in agreement with the result obtained in \cite{Chakraborty:2001wv} a consequence of Cosmic No Hair Conjecture.
\section{Harmonic Oscillator analysis and First Integral formulation}
We adopt a transformation related to the metric scalar of the hyper-surface given by
\begin{equation}
	\Lambda=\sqrt{det{(\eta_{ab})}}=\sqrt{\eta}
\end{equation}
Essentially $\Lambda$ is associated with the volume
of  hyper-surface and $\Lambda=0$ hints the singularity.
The dynamical evolution of $\Lambda$ is given by \cite{Poisson:2009pwt}
\begin{equation}
	\dfrac{1}{\sqrt{\eta}}\dfrac{d\sqrt{\eta}}{dt}=K
\end{equation} and hence,
\begin{equation}
	\dfrac{d\Lambda}{dt}=\Lambda K\label{eq34*}
\end{equation}
so that the RE (\ref{eq19}) can be expressed in the form
\begin{equation}
	\dfrac{d^{2}\Lambda}{dt^{2}}+\omega_{0}^{2}\Lambda=0\label{eq31**}
\end{equation}
where $\omega_{0}^{2}=-F_{c}$. Equation (\ref{eq31**}) resembles the differential equation of a classical real harmonic oscillator if and only if $F_{c}<0$. In \cite{Kar:2006ms}, \cite{Chakraborty:2023ork}, \cite{Chakraborty:2023lav}  it is found that if RE can be expressed in real harmonic oscillator form then it is inevitable to have focusing. In this case, negative sign of Focusing curvature $F_{c}$ indicates a real harmonic oscillator and facilitates focusing as seen in Case-III of the previous section. This formulation also shows that frequency of a real harmonic oscillator depends on the Focusing curvature or effective curvature and we name $\omega_{0}$ as ``\textbf{\textbf{Curvature frequency}}". Since $\omega_{0}^{2}=(M_{c}+^{(3)}R-~^{(4)}R)>0$ implies focusing, thus mean curvature and 3-space curvature assists (defies) convergence while 4-space curvature defies (assists) it depending on the sign of the curvatures. This may be interpreted in the way that since both $M_{c}$ and $^{(3)}R$ are extrinsic curvatures and $^{(4)}R$ is an intrinsic curvature, so extrinsic curvature always facilitates convergence while intrinsic curvature opposes it provided their contributions are all positive and a significant dominance of intrinsic curvature over extrinsic curvature is required to have a non-singular model.  So, this harmonic oscillator analysis gives us an insight about the role of extrinsic and intrinsic curvatures towards convergence or focusing. Now a first integral of (\ref{eq31**}) is given by
\begin{equation}
\dot{\Lambda}^{2}=\sqrt{\left(F(\Lambda)+\eta_{0}\right)}\label{eq36}
\end{equation} where
\begin{equation}
	F(\Lambda)=2\int \left(^{(4)}R-^{(3)}R-M_{c}\right)\Lambda d\Lambda\label{eq37}
\end{equation} and $\eta_{0}$ is a constant of integration. Using (\ref{eq34*}) and (\ref{eq36}), one gets the analytic solution of the RE as
\begin{equation}
	K=\pm\sqrt{(F(\Lambda)+\eta_{0})}\Lambda^{-1}\label{eq38}
\end{equation} where $F(\Lambda)$ is given by equation (\ref{eq37}). Suppose the integral $ 2\int \left(^{(4)}R-^{(3)}R-M_{c}\right)\Lambda d\Lambda$ is finite and bounded, then from equation (\ref{eq38}) we can say that at the singularity (characterized by the zero volume of the hyper-surface i.e, $\Lambda=0$) $K\rightarrow \pm \infty$, case of complete divergence or convergence arises. Thus, we are able to convert the first order RE into a second order differential equation (\ref{eq31**}). Usually, a second order differential equation is the Euler-Lagrange equation corresponding to a Lagrangian if it obeys all the Helmholtz conditions \cite{Davis:1929}-\cite{Nigam:2016}. So, there is a natural search for a Lagrangian corresponding to which the Euler-Lagrange's equation yields (\ref{eq31**}). All the Helmholtz conditions hold provided $F_{c}=\left(^{(4)}R-^{(3)}R-M_{c}\right)$ is a function of $\Lambda$ only. Thus, one may construct a Lagrangian of the form
\begin{equation}
	\mathcal{L}=\dfrac{1}{2}\dot{\Lambda}^{2}-V(\Lambda)
\end{equation}
with,
\begin{equation}
	\dfrac{\delta V(\Lambda)}{d\Lambda}=F_{c}(\Lambda)\Lambda
\end{equation} where $V(\Lambda)$ is the potential of the dynamical system corresponding to the congruence of time-like geodesic. So, a Lagrangian formulation (and hence a Hamiltonian formulation) is an important consequence of the first integral formulation which gives an insight that the classical potential is dependent on the Focusing curvature which is a combination of both extrinsic as well as intrinsic curvatures. 
\section{Conclusion}
The paper attempts to deduce a curvature form of the celebrated RE, which is somewhat unconventional from the way it is defined in literature . The motivation is to see the effect of curvature in focusing of a congruence of geodesics. Explicitly, the paper focuses on the notion of extrinsic and intrinsic curvature and how they influence focusing of a congruence of time-like geodesic orthogonal to space-like hyper-surface. Focusing theorem has been restated as a consequence of this curvature form of RE and certain criteria for convergence and possible avoidance of singularity have been determined using a notion of ``\textit{Focusing curvature}" or effective curvature $F_{c}$.  This focusing curvature is a combination of $^{(4)}R$, Ricci scalar (intrinsic curvature), $^{(3)}R$ three-space curvature (extrinsic) and mean curvature $M_{c}$ (extrinsic). It is found that, focusing occurs for $F_{c}\leq0$ while singularity is avoided if $F_{c}>0$. Further a nice analogy of mean curvature with SEC on matter has been found which shows that focusing is expected to occur in space-times with negative mean curvature or equivalently, there is no focusing (hence no singularity) in spaces with positive mean curvature. Using the curvature form of RE, focusing in inflationary era has been studied which is found to be in agreement to the work \cite{Chakraborty:2001wv}. Subsequently, with a suitable transformation related to the metric scalar of the hyper-surface the first order RE is converted to a second order differential equation analogous to that of a Harmonic oscillator and this analysis shows the explicit contribution of all the curvatures namely, extrinsic and intrinsic curvature towards convergence. The harmonic oscillator form of RE is popular in literature. It says that if RE corresponds to a real harmonic oscillator equation then focusing is inevitable. To see the effect of curvature on frequency of the oscillator, this transition of RE to harmonic oscillator form has been adopted in the present work. Based on the analysis, it is found that extrinsic curvature favors convergence while intrinsic curvature opposes it, provided they are positive in sign.  Moreover, another motivation to write the Harmonic oscillator form is that a first integral of the second order differential equation (the harmonic oscillator equation) gives the analytic solution of the RE. In this interpretation, definition of singularity with divergence of the expansion scalar has been revisited. Finally, a Lagrangian formulation has been carried out corresponding to which the second order differential equation (to which the first order RE was converted) becomes the Euler-Lagrange equation with the help of Helmholtz conditions provided the focusing curvature is a function of the transformed variable only.
	\section*{Acknowledgment}
 The authors are thankful to the anonymous reviewer for valuable and insightful comments that increased the quality and visibility of the work. M.C thanks University Grants Commission (UGC) for providing the Senior Research Fellowship (ID:211610035684/JOINT CSIR-UGC NET JUNE-2021). S.C thanks FIST program of DST, Department of Mathematics, JU (SR/FST/MS-II/2021/101(C)). The authors are thankful to Inter University Centre for Astronomy and Astrophysics (IUCAA), Pune, India for their warm hospitality and research facilities where this work was carried out under the ``Visiting Research Associates" program of S.C.


\begin{thebibliography}{50}
	\bibitem{Wald:1984rg}
	R.~M.~Wald,
	``General Relativity,''
	Chicago Univ. Pr., 1984,
	doi:10.7208/chicago/9780226870373.001.0001
	\bibitem{Weinberg:1972kfs}
	S.~Weinberg,
	``Gravitation and Cosmology: Principles and Applications of the General Theory of Relativity,''
	John Wiley and Sons, 1972,
	ISBN 978-0-471-92567-5, 978-0-471-92567-5
	\bibitem{Raychaudhuri:1953yv}
	A.~Raychaudhuri,
	Phys. Rev. \textbf{98}, 1123-1126 (1955)
	\bibitem{Kar:2006ms}
	S.~Kar and S.~SenGupta,
	Pramana \textbf{69}, 49 (2007)
	\bibitem{Penrose:1964wq}
	R.~Penrose,
	Phys. Rev. Lett. \textbf{14}, 57-59 (1965)
	\bibitem{Hawking:1970zqf}
	S.~W.~Hawking and R.~Penrose,
	Proc. Roy. Soc. Lond. A \textbf{314}, 529-548 (1970)
	\bibitem{Senovilla:2022vlr}
	J.~M.~M.~Senovilla,
	Gen. Rel. Grav. \textbf{54}, no.11, 151 (2022)
	\bibitem{Senovilla:2021pdg}
	J.~M.~M.~Senovilla,
	Phil. Trans. A. Math. Phys. Eng. Sci. \textbf{380}, no.2222, 20210174 (2022)
	\bibitem{Senovilla:2014gza}
	J.~M.~M.~Senovilla and D.~Garfinkle,
	Class. Quant. Grav. \textbf{32}, no.12, 124008 (2015)
	\bibitem{Landsman:2022hrn}
	K.~Landsman,
	Gen. Rel. Grav. \textbf{54}, no.10, 115 (2022)
	\bibitem{Chakraborty:2023ork}
	M.~Chakraborty, A.~Bose and S.~Chakraborty,
	Phys. Scripta \textbf{98}, no.2, 025007 (2023)
	\bibitem{Chakraborty:2023yyz}
	M.~Chakraborty and S.~Chakraborty,
	Class. Quant. Grav. \textbf{40}, no.15, 155010 (2023)
	\bibitem{Choudhury:2021zij}
	S.~G.~Choudhury, A.~Dasgupta and N.~Banerjee,
	Int. J. Geom. Meth. Mod. Phys. \textbf{18}, no.08, 2150115 (2021)
	\bibitem{Chakraborty:2023rgb}
	M.~Chakraborty and S.~Chakraborty,
	Annals Phys. \textbf{460}, 169577 (2024)
	\bibitem{Chakraborty:2023lav}
	M.~Chakraborty and S.~Chakraborty,
	Mod. Phys. Lett. A \textbf{38}, no.28n29, 2350129 (2023)
	\bibitem{Chakraborty:2024wty}
	M.~Chakraborty and S.~Chakraborty,
	Phys. Scripta \textbf{99}, no.4, 045203 (2024)
\bibitem{Chakraborty:2024khs}
S.~Chakraborty and M.~Chakraborty,
Int. J. Geom. Meth. Mod. Phys. \textbf{21}, no.10, 2440018 (2024)
	\bibitem{Chakraborty:2023voy}
	M.~Chakraborty and S.~Chakraborty,
	Annals Phys. \textbf{457}, 169403 (2023)
	\bibitem{Choudhury:2021huy}
	S.~G.~Choudhury, A.~Dasgupta and N.~Banerjee,
	Eur. Phys. J. C \textbf{81}, no.10, 906 (2021)
	\bibitem{Das:2013oda}
	S.~Das,
	Phys. Rev. D \textbf{89}, no.8, 084068 (2014)
	\bibitem{Blanchette:2020kkk}
	K.~Blanchette, S.~Das, S.~Hergott and S.~Rastgoo,
	Phys. Rev. D \textbf{103}, no.8, 084038 (2021)
	\bibitem{Dadhich:2005qr}
	N.~Dadhich,
	``Derivation of the Raychaudhuri equation,''
	[arXiv:gr-qc/0511123 [gr-qc]].
	\bibitem{Bhatt:2021hdi}
	R.~P.~Bhatt, A.~Roy and S.~Kar,
	Reson. \textbf{28}, no.3, 389-410 (2023)
	\bibitem{book1}
	O. Heckmann and E. Schucking, Z. Astrophysik 38, 95 (1955)
	\bibitem{book2}
	J. Ehlers, Contributions to the relativistic mechanics of continuous media, Gen. Rel. Grav.
	25, 1225 (1993) , English translation of original German article by P. Jordan, J. Ehlers, W.
	Kundt, R. K. Sachs, Proceedings of the Mathematical–Natural Sciences Section of the Mainz
	Academy of Sciences and Literature, Nr. 11, 792 (1961)
	\bibitem{Sachs:1962wk}
	R.~K.~Sachs,
	Proc. Roy. Soc. Lond. A \textbf{270}, 103-126 (1962)
	doi:10.1098/rspa.1962.0206
	\bibitem{book3}
	L. Landau and E. M. Lifshitz, Classical theory of fields, (Pergamom Press, Oxford, UK, 1975)
	\bibitem{Horwitz:2021lyc}
	L.~P.~Horwitz, V.~S.~Namboothiri, G.~Varma K, A.~Yahalom, Y.~Strauss and J.~Levitan,
	Symmetry \textbf{13}, no.6, 957 (2021)
	\bibitem{Burger:2018hpz}
	D.~J.~Burger, N.~Moynihan, S.~Das, S.~Shajidul Haque and B.~Underwood,
	Phys. Rev. D \textbf{98}, no.2, 024006 (2018)
	\bibitem{Bhattacharyya:2021djv}
	I.~Bhattacharyya and S.~Ray,
	Int. J. Mod. Phys. D \textbf{30}, no.12, 2150092 (2021)
	\bibitem{Albareti:2014dxa}
	F.~D.~Albareti, J.~A.~R.~Cembranos, A.~de la Cruz-Dombriz and A.~Dobado,
	JCAP \textbf{03}, 012 (2014)
	\bibitem{Albareti:2012se}
	F.~D.~Albareti, J.~A.~R.~Cembranos and A.~de la Cruz-Dombriz,
	JCAP \textbf{12}, 020 (2012)
	\bibitem{Poisson:2009pwt}
	E.~Poisson,
	``A Relativist's Toolkit: The Mathematics of Black-Hole Mechanics,''
	Cambridge University Press, 2009,
DOI:	\url{10.1017/CBO9780511606601}
	\bibitem{Chakraborty:2001wv}
	S.~Chakraborty and B.~C.~Paul,
	Phys. Rev. D \textbf{64}, 127502 (2001)
		\bibitem{Davis:1928}
	D.~R.~Davis,
	Trans. Amer. Math. Soc. \textbf{30} (1928), 710-736
	
	\bibitem{Davis:1929}
	D.~R.~Davis,
	Bull. Amer. Math. Soc. \textbf{35} (1929), 371-380
	
	\bibitem{Douglas:1941}
	J.~Douglas,
	Trans. Amer. Math. Soc. \textbf{50} (1941), 71-128
	
	\bibitem{Casetta:1941}
	L.~Casetta, C.~P.~Pesce
	Trans. Amer. Math. Soc. \textbf{50} (1941), 71-128
	
	\bibitem{Crampin:2010}
	M.~ Crampin, T.~ Mestdag and W.~ Sarlet
	Z. Angew. Math. Mech. \textbf{90} (2010), 502-508
	
	\bibitem{Nigam:2016}
	K.~ Nigam, K.~ Banerjee
	``A Brief Review of Helmholtz Conditions" 
	\url{arXiv:1602.01563}
	\end{thebibliography}
	\end{document}